Possible observation of energy level quantization in an intrinsic Josephson junction

H. Kashiwaya[a], T. Matsumoto[a], S. Kashiwaya[a], H. Shibata[a], H. Eisaki[a], Y. Yoshida[a], S. Kawabata[a], Y. Tanaka[b]

[a] National Institute of Advanced Industrial Science and Technology (AIST),
Umezono, Ibaraki, Tsukuba 305-8568, Japan

[b] Faculty of Science and Technology, Nagoya University, Nagoya, 464-8603, Japan

Abstract

Energy level quantization (ELQ) is studied to clarify the macroscopic quantum dynamics of the *d*-wave Josephson junction (JJ). The influences of the nodal quasiparticles of *d*-wave superconductivity on the damping effect are numerically evaluated on the basis of a phenomenological model. The calculation, based on realistic parameters for a $Bi_2Sr_2CaCu_2O_{8+\delta}$ (Bi2212) intrinsic JJ, shows that the observation of ELQ is possible when the sweep rate of the bias current exceeds 10 A/sec. High-sweep-rate measurements (121A/sec) performed on a Bi2212 intrinsic JJ result in the appearance of multiple peaks in the switching current distribution suggesting the realization of ELQ in the *d*-wave JJ.

1. Introduction

In recent decades, a wide variety of candidates for qubits have been presented. The superconducting qubit is believed to have the advantage that the decoherence due to the low-energy quasiparticles is largely suppressed because of the opening of the superconducting energy gap in the quasiparticle density of states [1]. Actually, various groups have reported the successful observation of quantum coherence in Josephson junctions (JJs) made of metal superconductors such as Al and Nb [2]. Recently, quantum coherence in high-$T_c$ superconductors (HTSCs) has attracted much attention. HTSCs have relatively large energy gaps and higher plasma frequency compared with those of metal superconductors, indicating their high potential for quantum electronic devices. On the other hand, JJs made of HTSCs have the drawback that the pairing symmetry is *d*-wave; thus, the quasiparticle current originating from the node of the pair potential inevitably has a dissipative component that may cause intrinsic damping in the dynamics of the JJ [3, 4]. Therefore, it is important to clarify the influence of the intrinsic damping on the switching dynamics of the *d*-wave JJ. In previous studies, observations of macroscopic quantum tunneling (MQT) revealed one aspect of the macroscopic quantum nature of *d*-wave JJs [5-8]. Here, we report the observation of energy level quantization (ELQ), which is another aspect of the macroscopic quantum nature of JJs, and we demonstrate the high potential of the intrinsic JJ for quantum electronics.

2. Energy level quantization and intrinsic damping

The dynamics of the JJ is described by the resistively shunted junction (RSJ) model. Fig. 1 shows the schematic diagrams of the washboard potential of the current-biased JJ. The gradient of the potential corresponds to the amplitude of the bias current (I). As the result of the quantum nature of the JJ, quantized energy levels are formed in the washboard potential when the bias current is smaller than the switching current ($I_{sw}$). Once the bias current exceeds $I_{sw}$, the junction switches to a finite-voltage state. If the sweep rate (dI/dt) of the bias current is low, the repopulation of the probability distribution inside the potential occurs at each switching event, whereas if dI/dt is sufficiently high (we will clarify this condition later), the repopulation of the probability distribution cannot follow the bias current variation; thus, the peaks corresponding to the quantized energy levels are directly observed in the switching current distribution (SCD) (see the P(I)-$I_{sw}$ diagram in Fig. 1). This ELQ has been observed for low-$T_c$ JJs [9,10].

The condition allowing the observation of ELQ strongly depends on the damping of the junction. Several papers have been presented that formulate the damping effects in the quantum dynamics of *d*-wave JJs [3,4]. Here, we phenomenologically estimate the influences of the nodal quasiparticles using realistic parameters for $Bi_2Sr_2CaCu_2O_{8+\delta}$ (Bi2212). In the above-mentioned RSJ model, the Q factor is given by $\omega_{pl}RC$, where $\omega_{pl}$ is the plasma frequency, R is the shunted resistance and C is the shunted capacitance. We assume that the effective resistance R of the Intrinsic Josephson Junction (IJJ) is simply given by $1/R=1/R_{qp}+1/R_s$, where $R_{qp}$ and $R_s$ are the resistances of the d-wave quasiparticles corresponding to the intrinsic damping and the unintentional microshorts due to the degradation of the insulator, respectively. In the case of *s*-wave superconductors, since $R_{qp}$ is small at sufficiently low temperatures, $R_s$ is the dominant component for the damping in a conventional metal superconductor JJ. However, in the case of a *d*-wave superconductor JJ, the resistance due to the intrinsic damping originating from the nodal quasiparticles ($R_{qp}$) is the dominant component if the junction quality is sufficiently high. We calculate $R_{qp}$ in the following way. A simple c-axis-oriented *d*-wave/insulator/*d*-wave (d/I/d) junction is assumed in the model, and the quasiparticle current I(eV) is evaluated as a function of the bias voltage V,

$$I(eV) = \frac{1}{eR_n} \int_{-\infty}^{\infty} N_d(E)N_d(E+eV)[f(E)-f(E+eV)]dE, \qquad (1)$$

where $N_d(E)$ is the density of states of the *d*-wave superconductor, f(E) is the Fermi distribution function and $R_n$ is the resistance when the junction is in the normal state. Although $N_d(E)$ is almost linear near the Fermi level, the resistance due to the quasiparticle current exhibits strong nonlinearity near the zero-bias level. For simplicity, we define the effective resistance as $R_{qp} \equiv I/V|(eV=h\omega_{pl})$. The calculated results of $R_{qp}$ and the Q factor by assuming realistic parameters for Bi2212 ($R_n$=100 Ω, Δ=40 meV or 50 meV, $\omega_{pl}$=65 GHz or 130 GHz, C=100 fF) are shown in Fig. 2. As is clear from the figure, $R_{qp}$ is sensitive to the amplitude of Δ. A Q factor of as high as 5,000-20,000 can be realized near 4 K owing to the relatively large Δ values of Bi2212 if $R_s$ is sufficiently high ($R_s = \infty$ is assumed).

According to the formula presented in Ref.[9], the ELQ can be clearly observed when

$$L \equiv \left[ R_{qp} C \frac{dI}{dt} \frac{1}{I_c} \right]^{-1} < 20. \qquad (2)$$

This condition is satisfied if we apply a bias current with a sweep rate higher than 10 A/sec by assuming that $I_c$ = 20 μA. These analyses indicate that the ELQ can be observed in the *d*-wave JJ if the damping is determined only by the nodal quasiparticles of the *d*-wave pair potential.

3. Experimental Results

The IJJs in this study were fabricated as follows. First, a single crystal of Bi2212 with a critical temperature ($T_c$) of 90 K was grown by the traveling solvent floating-zone method. Next, the crystal was cut into pieces of size $1000 \times 50 \times 10$ μm$^3$ by mechanical polishing and cleaving. The samples were glued onto SiO$_2$ substrates, and two Au contact electrodes were deposited onto each end to achieve electrical contacts. An S-shaped junction structure was formed by necking the center region and by forming slits from the side using a focused ion beam (FIB). The fabricated IJJ size was $1 \times 1 \times 0.01$ μm$^3$. The IJJs were quickly mounted on the sample holder of the probe and then slowly cooled down in a liquid helium Dewar. To obtain the SCD, typically 10,000 I-V curves were sampled using a digital oscilloscope.

Two SCDs obtained on a Bi2212 IJJ are shown in Fig. 3. The values of dI/dt for the applied bias voltage were about 14.1 A/sec and 121.4 A/sec, and both distributions were measured at 4.2 K. At lower values of dI/dt (not shown in the figure), the SCD has only one peak indicating thermal activation, consistent with the results of other studies. At the sweep rate of 14.1 A/sec, the SCD still has a single peak, as shown in Fig. 3(a). As the sweep rate increases further (121.4 A/sec), the distribution becomes wider and periodic peaks become clearly apparent in the SCD. This trend is consistent with that obtained in a previous study on Nb junctions[9]. We thus conclude that the observed SCD reflects the ELQ of the IJJ.

At present, only one junction has shown this type of response in our experiments. And we did not observe the ELQ with using other IJJs even at lower temperature of 1.7K. This is possibly because the extrinsic damping corresponding to $R_s$ in the above discussion exists in most of the actual junctions. In such a case, the temperature dependence of the Q factor is limited to a certain value corresponding to $R_s$ as schematically shown in Fig. 2(b); thus, the ELQ cannot be observed even in the mK region. We are now improving the sample fabrication process to confirm the origin of this sample dependence.

4. Conclusions

In this paper, the ELQ in d-wave JJs was investigated. The result suggests that a Bi2212 intrinsic JJ has a sufficiently high Q factor to observe the quantum nature in the switching dynamics. We believe that the observation of the ELQ demonstrates the high potential of HTSCs as superconducting qubits.

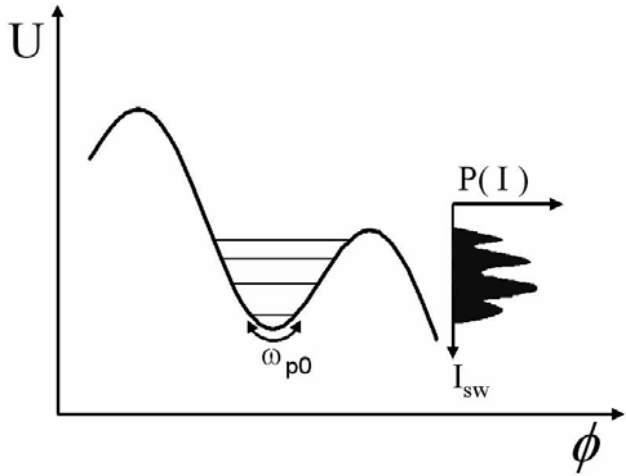

Fig. 1 Schematic diagrams of the washboard potential and quantized energy levels of the current-biased JJ. In the case that dI/dt is sufficiently high (L < 200), multiple peaks corresponding to the quantized energy levels are observed in the switching current distribution.

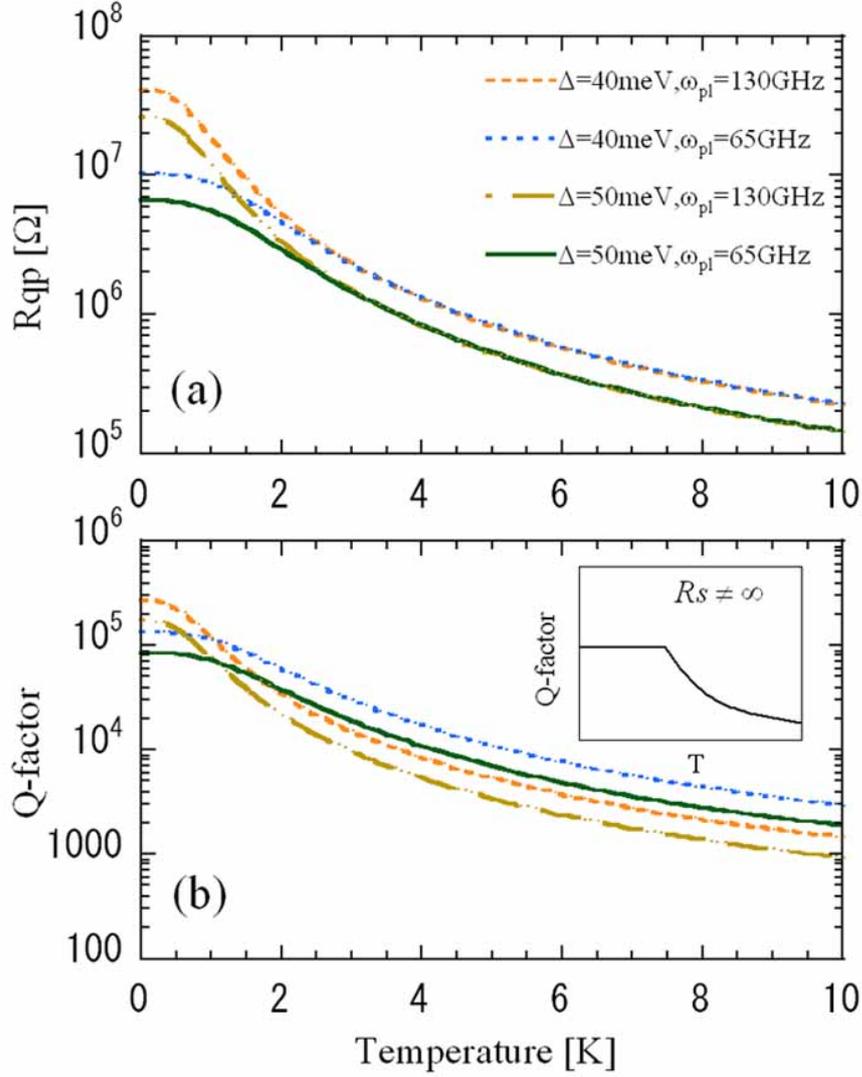

Fig. 2 Calculated temperature dependences of $R_{qp}$ and Q factor for *d*-wave IJJs by assuming realistic parameters for Bi2212 ($R_n$=100 Ω, Δ=40 meV or 50 meV, $\omega_{pl}$=65 GHz or 130 GHz, C=100 fF).

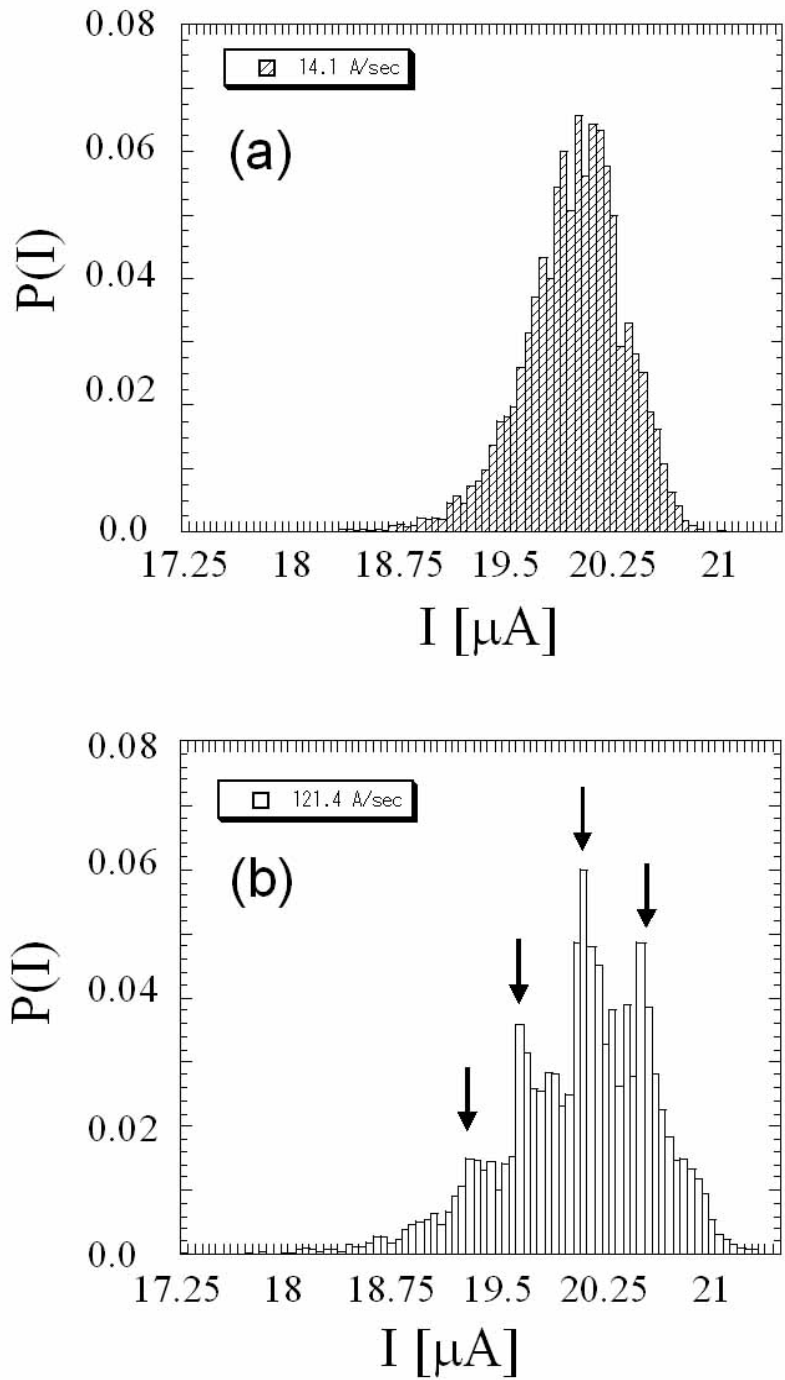

Fig. 3 Experimental data showing SCD for (a) dI/dt=14.1 A/sec (b) dI/dt=121.4 A/sec. Both distributions were measured at 4.2 K. The appearance of multiple peaks in the SCD indicates the quantization of the energy levels in the Bi2212 IJJ.